\documentclass[aps,prc,twocolumn,superscriptaddress,nofootinbib,floatfix]{revtex4-2}

\usepackage{graphicx}
\usepackage{amsmath}
\usepackage{amssymb}
\usepackage{bm}
\usepackage{dcolumn}
\usepackage{xcolor}
\graphicspath{{figures/}}

\newcommand{\gs}{\vert \Psi_0 \rangle}
\newcommand{\src}{\vert \Phi \rangle}
\newcommand{\lit}{\vert \tilde{\Psi}(\sigma) \rangle}

\begin{document}

%\title{Improved response function accuracy using alternative kernels\\ in the Lorentz %integral transform method}
\title{Improving the accuracy of the Lorentz Integral Transform method using 
complex kernels}

\author{Elad Parnes}
\affiliation{The Racah Institute of Physics, The Hebrew University of Jerusalem, Jerusalem 9190401, Israel}
\author{Nir Barnea}
\affiliation{The Racah Institute of Physics, The Hebrew University of Jerusalem, Jerusalem 9190401, Israel}

\date{\today}

\begin{abstract}
The Lorentz integral transform (LIT) method is a powerful tool for calculating quantum response functions; however, it requires an ill-posed inversion. 
Here we show that solutions of the LIT equation can also determine two additional 
integral transforms: a complex Stieltjes transform and a double-pole transform, 
at little additional computational cost. A Fourier analysis of the associated 
deconvolution problem shows that both alternative kernels are better conditioned than the Lorentzian kernel at the same width parameter $\Gamma $. We benchmark the resulting inversions for deuteron photodisintegration by adding controlled noise to the LIT solutions. Relative to the standard LIT, the alternative kernels yield response functions with reduced noise-induced scatter: by a factor of 2{--}3 for the complex Stieltjes kernel and 3{--}4 for the double-pole kernel.

\end{abstract}

\maketitle

\section{Introduction}
\label{sec:intro}

Response functions are central objects of quantum many-body theory. They
encode the way a system reacts to an external probe and thereby connect microscopic
structure and dynamics to scattering and reaction experiments in nuclear, atomic, and
condensed-matter physics~\cite{FetterWalecka,Mahan}. With nuclear physics as
our prime interest: inclusive electron scattering cross sections, photoabsorption, and
neutrino-nucleus scattering are all governed by the nuclear response
functions~\cite{Bacca:2014tla}. Computing an inclusive response $R(\omega)$ from first
principles is, however, notoriously difficult task, primarily because it requires a sum 
(integral) over \emph{all}
final states of the system, including the many-body continuum.
Beyond the lightest systems, an explicit construction of continuum
wavefunctions is out of reach. The Lorentz integral transform (LIT)
method~\cite{LIT,LITReview} circumvents this obstacle: instead of $R(\omega)$ itself, one
computes its convolution with a Lorentzian kernel,
\begin{equation}
  L(\sigma,\Gamma)=\int d\omega\,
  \frac{R(\omega)}{(\omega-\sigma)^{2}+\Gamma^{2}}\,,
  \label{eq:LITdef}
\end{equation}
which, as recalled below, is obtained from the solution of a bound-state-like equation
with standard few- and many-body techniques. The LIT method has enabled \emph{ab initio}
calculations of electromagnetic and electroweak responses from the few-nucleon sector up
to medium-mass nuclei and nuclear
matter~\cite{Bacca13,Lovato:2013cua,Birkhan:2016qkr,Sobczyk:2021dwm,Sobczyk:2023sxh,Acharya:2024xah,Sobczyk:2024hdl},
and has recently been combined with neural-network quantum states~\cite{Parnes:2026},
promising a further increase in the span of reachable systems.

The price of the approach is that the physical response must be recovered from
$L(\sigma,\Gamma)$ by inverting Eq.~\eqref{eq:LITdef}, a classic ill-posed problem:
features of $R(\omega)$ narrower than $\Gamma$ are exponentially suppressed in the
transform, and their reconstruction amplifies any error in $L$ accordingly. In practice
the inversion is manageable, but keeping it under control requires 
$\Gamma$ to be similar to the width of $R(\omega)$ or smaller. 
This creates a trade-off, since the smaller $\Gamma$ is, the
harder the LIT equation becomes: the spatial extent of its solution increases as the
complex energy $\sigma+i\Gamma$ approaches the continuum
spectrum, demanding larger model spaces~\cite{LITReview,LITINversionBarnea:2010}. It is
this trade-off, rather than the inversion per se, that limits the resolution of LIT
calculations and their extension to heavier systems.

The inversion step itself has received considerable attention. The standard procedure
expands $R(\omega)$ in a small set of basis functions fitted to the
transform~\cite{LITReview,ResponseLITEfros:2010}, and several regularized generalizations
have been studied systematically with respect to basis choice and regularization
strategy~\cite{Andreasi:2005rc,LITINversionBarnea:2010}. Alternatives include the maximum
entropy method~\cite{Skilling1989,Jarrell:1996rrw,Murata:2016vsl,Parnes:2026} and modern
Bayesian reconstructions~\cite{Burnier:2013nla,Rothkopf:2016luz}. All of these efforts, however,
address how to invert a \emph{given} transform. The complementary question -- whether the
kernel itself can be chosen so as to make the inversion better conditioned -- has
remained essentially unexplored; the Stieltjes kernel is mentioned in
Ref.~\cite{LITReview} as a possible alternative, but kernel choice has not been pursued
as a route to inversion stability. In this paper we show that two complex-valued
kernels, a complex Stieltjes kernel and a double-pole kernel, can be evaluated from the
LIT wavefunctions that any LIT calculation produces, at essentially no
extra cost, and that they yield substantially more stable inversions. We first derive
the transforms and explain the improvement analytically through a Fourier analysis of
the deconvolution problem, and then quantify it numerically in a deuteron
photodisintegration benchmark with controlled noise.

\section{Method}
\label{sec:method}

\subsection{Integral transforms}
\label{sec:general}

Consider a system with Hamiltonian $\hat{H}$, ground state $\gs$ of energy $E_0$, and an
excitation operator $\hat{O}$ representing the coupling to an external probe. The
inclusive response function is 
\begin{equation}
  R(\omega)=\sum_{f}\,
  \bigl|\langle\Psi_f|\hat{O}\,\gs\bigr|^{2}\,
  \delta\!\left(E_f-E_0-\omega\right),
  \label{eq:response}
\end{equation}
where the sum runs over all final states $|\Psi_f\rangle$ of energy $E_f$, with an
integration implied over the continuum part of the spectrum. A general integral 
transform of the response with kernel
$K(\omega,\bm{\alpha})$, depending on a set of parameters $\bm{\alpha}$, is
\begin{equation}
  L(\bm{\alpha})=\int d\omega\, K(\omega,\bm{\alpha})\,R(\omega)\,.
  \label{eq:generalIT}
\end{equation}
Substituting Eq.~\eqref{eq:response} into Eq.~\eqref{eq:generalIT}, carrying out the
$\omega$ integration against the delta function, and using the completeness of the
energy eigenstates, $\sum_f |\Psi_f\rangle\langle\Psi_f| = \hat{1}$, one obtains the
key identity~\cite{LITReview}
\begin{equation}
  L(\bm{\alpha})
  =\langle\Phi|K(\hat{H}-E_0,\bm{\alpha})\src\,,
  \label{eq:keyidentity}
\end{equation}
where we introduced the source state $\src\equiv\hat{O}\gs$. Equation
\eqref{eq:keyidentity} expresses the transform as a ground-state expectation value:
no information about individual final states is needed. In particular, 
for rational kernels with a denominator of degree at most two and
with poles at complex energies,
$K(\hat H - E_0)$ is built from resolvents $(\hat{H}-E_0-z)^{-1}$ with
$\operatorname{Im}z\neq 0$; each resolvent acting on the source defines an auxiliary
state through a Schr\"odinger-like equation with a complex energy shift. Because a
Hermitian $\hat H$ has no normalizable eigenstate at complex energy, this equation has a
unique, spatially localized solution, which is the property that makes the whole
approach practical with bound-state methods. This observation motivates the central question of this work: among
the kernels accessible in this way, which one leads to the most stable inversion?

\subsection{The Lorentz integral transform}
\label{sec:LIT}

The LIT method employs the Lorentzian kernel
\begin{align}
  K_{\mathrm{LIT}}(\omega;\sigma,\Gamma)
  &=\frac{1}{(\omega-\sigma)^{2}+\Gamma^{2}}
  \nonumber\\
  &=\frac{1}{(\omega-\sigma-i\Gamma)(\omega-\sigma+i\Gamma)}\,,
  \label{eq:LITkernel}
\end{align}
a bell-shaped function of width $\Gamma$ centered at the parameter $\sigma$, which is
scanned across the energy range of interest. Factorizing the kernel as in
Eq.~\eqref{eq:LITkernel} and applying Eq.~\eqref{eq:keyidentity}, the transform becomes
the squared norm of a single auxiliary state,
\begin{equation}
  L_{\mathrm{LIT}}(\sigma,\Gamma)
  =\langle\tilde{\Psi}(\sigma)\vert\tilde{\Psi}(\sigma)\rangle\,,
  \label{eq:LITnorm}
\end{equation}
where the LIT wavefunction $\lit$ solves the LIT equation
\begin{equation}
  \left(\hat{H}-E_0-\sigma+i\Gamma\right)\lit=\src\,.
  \label{eq:LITeq}
\end{equation}
Equation~\eqref{eq:LITeq} is solved once per value of $\sigma$ (at fixed $\Gamma$), using
whichever bound-state method suits the system at hand. As $\Gamma\to 0$ at fixed
$\sigma$ inside the continuum, the solution of Eq.~\eqref{eq:LITeq} approaches a
scattering-like state: its spatial extent grows and the numerical effort with it. This is
the quantitative origin of the trade-off described in the Introduction, and the reason
one would like to extract as much information as possible from calculations performed at
moderate values of $\Gamma$.

\subsection{Alternative kernels from the same wavefunctions}
\label{sec:kernels}

The solutions $\lit$ of Eq.~\eqref{eq:LITeq} contain more information than the norm in
Eq.~\eqref{eq:LITnorm} reveals. We consider two additional kernels, both evaluated
directly from the LIT states $\lit$ and source $\src$.

\begin{figure}[t]
\includegraphics[width=\columnwidth]{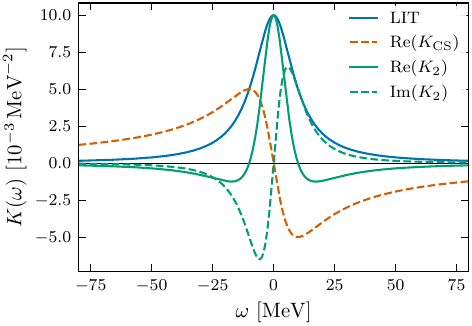}
\caption{\label{fig:kernels}%
The integral-transform kernels of Table~\ref{tab:kernels} as functions of $\omega$ (at
$\sigma=0$) for $\Gamma=10$~MeV. The standard LIT kernel (solid blue) is symmetric,
positive, and coincides with the imaginary part of the complex Stieltjes kernel. The
real part of the complex Stieltjes kernel (dashed red) is antisymmetric and carries
sign information absent from the LIT. The double-pole kernel $K_2$ has a real part
(solid green) with the same peak height $1/\Gamma^2$ but a narrower central structure
and negative side lobes, and an antisymmetric imaginary part (dashed green). The chosen
normalizations make all kernels directly comparable in magnitude.}
\end{figure}

The first is a \emph{complex Stieltjes} kernel,
\begin{equation}
  K_{\mathrm{CS}}(\omega;\sigma,\Gamma)
  =-\frac{1}{\Gamma}\,\frac{1}{\omega-\sigma+i\Gamma}\,,
  \label{eq:CSkernel}
\end{equation}
a Stieltjes-type kernel~\cite{LITReview} with its pole moved off the real axis by
$\Gamma$. By Eq.~\eqref{eq:keyidentity} its transform is a single overlap with the
source,
\begin{equation}
  L_{\mathrm{CS}}(\sigma,\Gamma)
  =-\frac{1}{\Gamma}\,\langle\Phi\vert\tilde{\Psi}(\sigma)\rangle\,.
  \label{eq:CStransform}
\end{equation}
The normalization in Eq.~\eqref{eq:CSkernel} is chosen such
that the imaginary part of the kernel reproduces the Lorentzian exactly,
\begin{equation}
  \operatorname{Im}K_{\mathrm{CS}}=K_{\mathrm{LIT}}\,,
  \qquad
  \operatorname{Re}K_{\mathrm{CS}}=-\frac{\omega-\sigma}{\Gamma}\,K_{\mathrm{LIT}}\,,
  \label{eq:CSparts}
\end{equation}
so that the complex Stieltjes transform contains the standard LIT as its imaginary part,
supplemented by an antisymmetric real part that carries independent, sign-sensitive
information about the response, see Fig.~\ref{fig:kernels}.
Equation~\eqref{eq:CSparts} also defines a useful
intermediate object: the imaginary part of Eq.~\eqref{eq:CStransform} is an alternative
estimator of the \emph{standard} LIT,
\begin{equation}
  L_{\mathrm{GS}}(\sigma,\Gamma)
  =\operatorname{Im}L_{\mathrm{CS}}(\sigma,\Gamma)
  =-\frac{1}{\Gamma}\operatorname{Im}\langle\Phi\vert\tilde{\Psi}(\sigma)\rangle\,,
  \label{eq:GS}
\end{equation}
which we denote LIT-GS. We stress that this estimator is not new: it was already
employed in the neural-network LIT calculations of Ref.~\cite{Parnes:2026}, where the
overlap with the source proved advantageous for stochastic solvers.
Analytically $L_{\mathrm{GS}}=L_{\mathrm{LIT}}$, but the two
estimators respond differently to imperfections in $\lit$: the norm
\eqref{eq:LITnorm} is quadratic in the solution, so uncorrelated errors do not average
out and instead produce a positive bias, whereas the overlap \eqref{eq:CStransform} is linear in
$\lit$ and projects the error onto the smooth, fixed source state. The LIT-GS estimator
thus isolates the part of the improvement that stems from error propagation alone,
independent of the kernel shape.

The second kernel is obtained by squaring the resolvent, giving a \emph{double pole} at
$\omega=\sigma-i\Gamma$; equivalently, it is a width derivative of the underlying
Stieltjes kernel:
\begin{equation}
  K_{2}(\omega;\sigma,\Gamma)
  =-\frac{1}{(\omega-\sigma+i\Gamma)^{2}}
  =i\,\frac{\partial}{\partial\Gamma}\!\left[\Gamma\,K_{\mathrm{CS}}\right].
  \label{eq:K2kernel}
\end{equation}
Its real part is a peak of height $1/\Gamma^{2}$ at $\omega=\sigma$, the same height as
the Lorentzian but with a visibly narrower core and negative side lobes, while its
imaginary part is antisymmetric, see Fig.~\ref{fig:kernels}. The corresponding
transform is
\begin{equation}
  L_{2}(\sigma,\Gamma)
  =-\langle\Phi\vert
  \left(\hat{H}-E_0-\sigma+i\Gamma\right)^{-2}\src\,.
  \label{eq:K2transform}
\end{equation}
Evaluating this expression requires, in addition to the regular solution
$\vert\tilde{\Psi}(\sigma,\Gamma)\rangle$ of Eq.~\eqref{eq:LITeq} (with the dependence
on $\Gamma$ usually kept implicit for brevity), the solution
$\vert\tilde{\Psi}(\sigma,-\Gamma)\rangle$ obtained with the opposite sign of
$i\Gamma$, since
\begin{equation}
  L_{2}(\sigma,\Gamma)
  =-\langle\tilde{\Psi}(\sigma,-\Gamma)\vert\tilde{\Psi}(\sigma,\Gamma)\rangle\,.
  \label{eq:K2bracket}
\end{equation}
In the worst case this requires solving the LIT equation twice, but in many cases of interest it will not be necessary. For a time-reversal-invariant Hamiltonian, $[\hat{H},\hat{T}]=0$, the two solutions are
related by
\begin{align}
  \vert\tilde{\Psi}(\sigma,-\Gamma)\rangle
  &=\hat{T}^{2}\left(\hat{H}-E_0-\sigma-i\Gamma\right)^{-1}\src
  \nonumber\\
  &=\hat{T}\left(\hat{H}-E_0-\sigma+i\Gamma\right)^{-1}\hat{T}\src
  \nonumber\\
  &=\hat{T}\,\vert\tilde{\Psi}(\sigma,\Gamma)\rangle\,,
  \label{eq:Trelation}
\end{align}
where the antilinearity of $\hat{T}$ flips the sign of $i\Gamma$, and we assumed
$\hat{T}\src=\src$. The latter holds for a time-reversal-invariant excitation 
operator acting on a
time-reversal-invariant ground state. For a non-degenerate ground state this property is guaranteed. In the degenerate case it can be enforced by projecting
$\gs\to\tfrac{1}{2}(1\pm\hat{T})\gs$, choosing a sign for which the projection is
nonzero and adjusting for that sign in the final formula. Furthermore, if $[\hat{H},\hat{K}]=0$, with $\hat{K}$ the
complex-conjugation operator -- the case, for example, for effective nuclear potentials
without spin-orbit terms -- one obtains the even simpler relation
$\vert\tilde{\Psi}(\sigma,-\Gamma)\rangle=\hat{K}\vert\tilde{\Psi}(\sigma,\Gamma)\rangle$
(with $\src$ likewise chosen real), and the transform reduces to 
\begin{equation}
  L_{2}(\sigma,\Gamma)
  =-\langle\tilde{\Psi}^{*}(\sigma,\Gamma)\vert\tilde{\Psi}(\sigma,\Gamma)\rangle\,,
  \label{eq:K2unconj}
\end{equation}
where $\vert\tilde{\Psi}^{*}\rangle$ denotes the complex conjugate of the
coordinate-space wavefunction.

Table~\ref{tab:kernels} summarizes the kernels, the matrix elements from which
they are computed, and their Fourier transforms, to which we turn next.

\begin{table*}[t]
\caption{\label{tab:kernels}%
The integral-transform kernels considered in this work. The last column lists the
Fourier transform of the kernel
$K(\omega-\sigma,\Gamma)$, at $\sigma=0$, entering the convolution representation of
the transform, Eq.~\eqref{eq:convolution}; $\Theta(k)$ denotes the Heaviside step
function.}
\begin{ruledtabular}
\begin{tabular}{llll}
Kernel & $K(\omega;\sigma,\Gamma)$ & Computed as & $\widetilde{K}(k)$ \\
\hline
Standard LIT & $\dfrac{1}{(\omega-\sigma)^{2}+\Gamma^{2}}$ &
$\langle\tilde{\Psi}(\sigma)\vert\tilde{\Psi}(\sigma)\rangle$ &
$\dfrac{\pi}{\Gamma}\,e^{-\Gamma|k|}$ \\[2.2ex]
LIT-GS & $\dfrac{1}{(\omega-\sigma)^{2}+\Gamma^{2}}$ &
$-\dfrac{1}{\Gamma}\operatorname{Im}\langle\Phi\vert\tilde{\Psi}(\sigma)\rangle$ &
$\dfrac{\pi}{\Gamma}\,e^{-\Gamma|k|}$ \\[2.2ex]
Complex Stieltjes & $-\dfrac{1}{\Gamma}\,\dfrac{1}{\omega-\sigma+i\Gamma}$ &
$-\dfrac{1}{\Gamma}\langle\Phi\vert\tilde{\Psi}(\sigma)\rangle$ &
$\dfrac{2\pi i}{\Gamma}\,e^{-\Gamma k}\,\Theta(k)$ \\[2.2ex]
Double-pole ($K_2$) & $-\dfrac{1}{(\omega-\sigma+i\Gamma)^{2}}$ &
$-\langle\tilde{\Psi}^{*}(\sigma)\vert\tilde{\Psi}(\sigma)\rangle$ &
$2\pi k\,e^{-\Gamma k}\,\Theta(k)$ \\
\end{tabular}
\end{ruledtabular}
\end{table*}

\subsection{Fourier analysis of inversion stability}
\label{sec:fourier}

We note that all the kernels above depend on $\omega$ and $\sigma$ only through the 
difference $\omega-\sigma$, i.e. $K(\omega,\sigma,\Gamma) = K(\sigma-\omega;\Gamma)$.
Hence, their integral-transform is a convolution,
\begin{equation}  \label{eq:convolution}
  L(\sigma)=\int d\omega\,K(\sigma-\omega)\,R(\omega)
  =\left(K*R\right)(\sigma)\,.
\end{equation}
Considering the Fourier transform $\mathcal{F}$ with normalization
$\widetilde{f}(k)=\mathcal{F}[f](k)=\int d\omega\,f(\omega)\,e^{-ik\omega}$, the convolution theorem turns
Eq.~\eqref{eq:convolution} into a product, and the formal inversion of the 
integral-transform reads
\begin{equation}
  R(\omega)=\mathcal{F}^{-1}\!\left[
  \frac{\widetilde{L}(k)}{\widetilde{K}(k)}\right](\omega)\,.
  \label{eq:fourierinversion}
\end{equation}
In this form the ill-posedness is transparent: Fourier components $k$ of the error in
$\widetilde{L}$ are amplified in the reconstruction by $1/|\widetilde{K}(k)|$. The
stability of the deconvolution is therefore governed by how slowly
$|\widetilde{K}(k)|$ decays at large $k$.

The Fourier transforms of the kernels are elementary and are listed in the last column
of Table~\ref{tab:kernels}. All decay with the same exponential rate
$e^{-\Gamma k}$; this rate is set by the distance $\Gamma$ of the kernel poles from the
real axis and is the fundamental resolution scale shared by all transforms computed from
the same LIT solutions. The differences between the kernels are nevertheless
substantial. Consider first the complex Stieltjes kernel against the Lorentzian. Because
all its poles lie on the same side of the real $\omega$ axis,
$\widetilde{K}_{\mathrm{CS}}(k)$ vanishes identically for $k<0$: the Fourier support is
\emph{one-sided}. Such a kernel lets more information through the transform: the
standard LIT is obtained locally from the complex Stieltjes transform by taking the
imaginary part, Eq.~\eqref{eq:GS}, whereas the converse reconstruction is nonlocal and
noise amplifying. Note that no information is lost on the missing half-axis, because the
response function is real and thus $\widetilde{R}(-k)=\widetilde{R}^{*}(k)$; were the
response not real, the one-sided kernel would render the inversion impossible. Comparing
next the double-pole kernel with the complex Stieltjes, their Fourier transforms differ
by a factor proportional to $k$. At large $k$ the double-pole kernel decays more slowly,
yielding better stability precisely where the deconvolution is most delicate. Together
these features predict the stability ordering
\begin{equation}
  K_2 \;\succ\; K_{\mathrm{CS}} \;\succ\; K_{\mathrm{LIT}}\,,
  \label{eq:ordering}
\end{equation}
which the numerical experiments of Sec.~\ref{sec:results} confirm.

One caveat is in order. The factor $k$ in $\widetilde{K}_2$ suppresses the low-$k$
region: the double-pole transform is insensitive to the addition of a constant to the
response function and, more generally, to contributions varying on wavelengths much
larger than $\Gamma$. This could become problematic if such long-wavelength components
are not otherwise suppressed by the ansatz adopted for the response. When the low-$k$
content is important, the issue can be circumvented by a hybrid approach in which the
complex Stieltjes and double-pole transforms are computed and inverted together, the
former being maximal at $k=0$. We tried such a hybrid scheme for the benchmark presented
below and obtained no more than a few percent improvement in the inversion stability.

\subsection{Inversion procedure}
\label{sec:inversion}

For the numerical inversion we follow the regularized basis-expansion procedure of
Refs.~\cite{LITReview,Andreasi:2005rc,Parnes:2026}, to which we refer for a more
extended discussion. The response is expanded as
\begin{equation}
  R(\omega)\;\approx\;\sum_{n=1}^{N_{\max}}c_{n}\,
  \chi_{n}(\omega;\alpha_1,\alpha_2)\,,
  \label{eq:ansatz}
\end{equation}
with
\begin{equation}
  \chi_{1}=\omega'^{\,\alpha_1}e^{-\alpha_2\omega'}\,,\quad
  \chi_{n\ge2}=\omega'^{\,\alpha_1}
  \exp\!\!\left(-\frac{\alpha_2\,\omega'}{n\,\beta}\right),
  \label{eq:basis}
\end{equation}
where $\omega'=\omega-\omega_{\mathrm{th}}$ and $\omega_{\mathrm{th}}$ is the threshold
energy, while $\alpha_1$, $\alpha_2$, and $\beta$ are nonlinear shape parameters. Each
$\chi_n$ is normalized to unit $L^2$ norm. For a given $(\alpha_1,\alpha_2,\beta)$, the
transform of each
basis function is computed by direct quadrature of Eq.~\eqref{eq:generalIT} with the
appropriate kernel, and the linear coefficients $c_n$ are obtained from a least-squares
fit of $\sum_n c_n L[\chi_n](\sigma)$ to the transform data over the $\sigma$ grid, with
$L^2$ regularization of the coefficients. For the complex transforms, real and imaginary
parts are fitted simultaneously. This procedure treats all kernels on an equal footing:
the only difference between inverting the standard LIT and inverting the alternative
transforms is the kernel used to map the basis functions onto the data.

\section{Results}
\label{sec:results}

\subsection{Deuteron benchmark}
\label{sec:benchmark}

We benchmark the four transforms -- standard LIT, LIT-GS, complex Stieltjes, and double
pole -- on the unretarded dipole response of the deuteron, i.e., deuteron
photodisintegration. This choice is deliberate: the two-body problem can be solved to
essentially arbitrary numerical precision, so the exact LIT wavefunctions are available
and every difference between inversions can be attributed to the transform itself rather
than to an imperfect many-body solution.

For the nuclear interaction we use the pionless effective-field-theory potential
``model o'' of Ref.~\cite{Schiavilla:2021dun}, a two-Gaussian regularized $S$-wave
interaction, which binds the deuteron in the $^{3}S_{1}$ channel with
$E_{0}=-2.242$~MeV. The source $\src$ is built by acting with the dipole operator on the
ground state, exciting the $^{3}P$ continuum; for this potential the odd partial waves
are non-interacting, so the final-state Hamiltonian contains only the kinetic
term. Both the ground state and the LIT equation~\eqref{eq:LITeq} are
solved as radial differential equations on a uniform grid of 800 points extending to
$r=80$~fm, the latter by a direct banded-matrix solution of the finite-difference
equations. The LIT wavefunctions are computed for $\Gamma=10$~MeV on a grid of
$\sigma\in[-100,200]$~MeV with $0.5$~MeV spacing, and all four transforms are then
extracted from the same set of solutions via
Eqs.~\eqref{eq:LITnorm},~\eqref{eq:CStransform},~\eqref{eq:GS},
and~\eqref{eq:K2unconj}, since the potential contains no spin-orbit term.

\subsection{Noise model and inversion comparison}
\label{sec:noise}

In realistic applications the LIT wavefunctions are never exact: they carry systematic
truncation errors and, in Monte Carlo or neural-network-based
approaches~\cite{Parnes:2026}, stochastic noise. To emulate this in a controlled way we
perturb the exact solutions with multiplicative complex Gaussian noise,
\begin{equation}
  \tilde{\Psi}(r_j,\sigma)\;\to\;
  \tilde{\Psi}(r_j,\sigma)\left[1+\epsilon\,\eta_j+i\,\epsilon\,\eta'_j\right],
  \label{eq:noisemodel}
\end{equation}
with $\eta_j,\eta'_j$ independent standard normal variables drawn at every grid point
$r_j$ and every $\sigma$. For each
$\epsilon\in\{0.01,0.03,0.05,0.08,0.1,0.15,0.2,0.4,0.5,0.7\}$ we generate $100$
independent noise realizations; all four transforms are evaluated from the same noisy
wavefunctions, so the comparison isolates how each kernel propagates identical input
errors. Each noisy transform is then inverted independently with
$N_{\max}=13$ basis functions and $\alpha_1=1$ fixed by the known threshold behavior of
the dipole response. The remaining nonlinear parameters $\alpha_2$ and $\beta$ were
optimized once on the noiseless transform and rounded to $\alpha_2=0.8$, $\beta=1.3$;
they were then held fixed across all noise realizations, reducing the inversion to a
purely linear problem and eliminating the additional noise that a per-realization
nonlinear optimization would introduce.

\begin{figure}[t]
\includegraphics[width=\columnwidth]{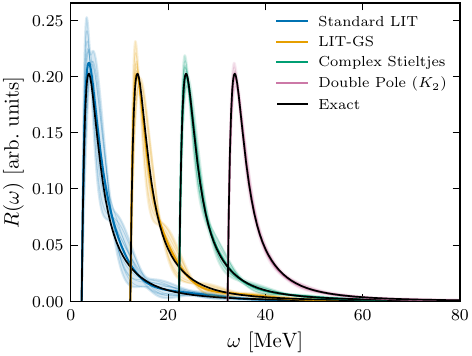}
\caption{\label{fig:inversion}%
Inverted response functions from 10 independent noise realizations at noise level
$\epsilon=0.2$, for the standard LIT (blue, natural position), the LIT-GS estimator
(orange, shifted by $+10$~MeV for clarity), the complex Stieltjes transform (green,
$+20$~MeV), and the double-pole transform $K_2$ (purple, $+30$~MeV). For each transform
the thin lines show the individual realizations, the thick line their mean, and the
shaded band the $\pm2\sigma$ spread; the black curve is the inversion of the
corresponding noiseless transform, which reproduces the exact response. The spread
shrinks visibly from left to right, with the double-pole reconstruction tightest around
the exact result.}
\end{figure}

Figure~\ref{fig:inversion} shows the outcome at noise level $\epsilon=0.2$ for the four
transforms. In all cases the ensemble mean remains close to the noiseless curve, but the
size of the scatter differs markedly: the standard LIT realizations fluctuate visibly in
both the peak height and the tail, the LIT-GS and complex Stieltjes reconstructions are
progressively tighter, and the double-pole reconstructions form a tight bundle around
the exact response. The overlap-based transforms are moreover
strictly unbiased under the noise model~\eqref{eq:noisemodel}, whereas the
norm~\eqref{eq:LITnorm} acquires a small positive bias of relative size
$2\epsilon^{2}$.

\subsection{Quantitative noise robustness}
\label{sec:robustness}

\begin{figure}[t]
\includegraphics[width=\columnwidth]{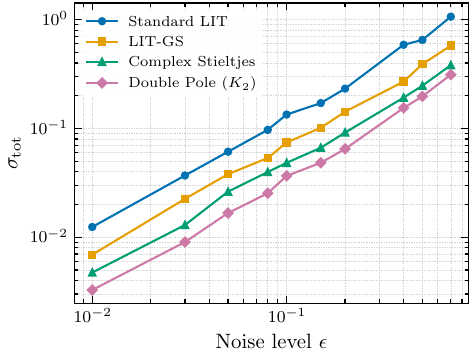}
\caption{\label{fig:noise}%
Integrated standard deviation $\sigma_{\mathrm{tot}}$ of the inverted response,
Eq.~\eqref{eq:sigmatot}, as a function of the noise level $\epsilon$, for the four
transforms considered ($100$ realizations per point; both axes logarithmic). The
ordering is the same at every noise level: the double-pole kernel is the most stable,
followed by the complex Stieltjes, the LIT-GS estimator, and the standard LIT.}
\end{figure}

To quantify the spread we use the integrated standard deviation of the reconstructed
response across the $100$ realizations,
\begin{equation}
  \sigma_{\mathrm{tot}}
  =\left(\sum_{i}\operatorname{Var}\!\left[R_{\mathrm{inv}}(\omega_i)\right]
  \right)^{1/2},
  \label{eq:sigmatot}
\end{equation}
evaluated on a uniform grid of $5000$ points $\omega_i$ spanning threshold to
$400$~MeV. Figure~\ref{fig:noise} displays $\sigma_{\mathrm{tot}}$ as a function of the
noise level for all four transforms.

Three features stand out. First, all four curves grow essentially linearly with
$\epsilon$ over nearly two orders of magnitude, so the comparison is meaningful at any
noise level. Second, the ordering is strict and uniform, and matches the prediction of
Eq.~\eqref{eq:ordering}: at every $\epsilon$ the double-pole inversion has the smallest
spread, a factor of $3.3$--$4.1$ below the standard LIT, with a factor of
$3.7$ at $\epsilon=0.1$, while the complex Stieltjes improves on the standard LIT by a
factor of $2.3$--$3.1$. Third, the LIT-GS estimator, which shares the Lorentzian kernel
with the standard LIT and differs only in being computed from the source overlap rather
than the norm, is more stable than the standard LIT by a factor of $1.6$--$2.2$. The
gains therefore decompose naturally into two mechanisms of comparable importance: the
linear-in-$\tilde\Psi$ error propagation of the overlap-based evaluations
(LIT~$\to$~LIT-GS), and the better-conditioned kernels acting on top of it
(LIT-GS~$\to$~CS~$\to$~$K_2$), in line with the Fourier analysis of
Sec.~\ref{sec:fourier}.

\section{Summary and outlook}
\label{sec:summary}

We have shown that the wavefunctions computed in any Lorentz-integral-transform
calculation determine, in addition to the standard LIT, two better-conditioned integral
transforms of the response function: a complex Stieltjes transform, obtained from the
overlap of the LIT solutions with the source, and a double-pole transform, obtained
through time-reversal symmetry from the bracket of the solutions with themselves. 
They require at most one additional solution of the LIT
equation.

A Fourier analysis of the deconvolution problem explains why these transforms invert
more stably. All kernels computed from the same LIT solutions share the exponential
spectral decay $e^{-\Gamma k}$, but the double-pole kernel gains a factor of $k$, and
the complex kernels concentrate their support on half the spectral axis while providing
two real data values per point, each carrying distinct information about the response.
In a deuteron photodisintegration benchmark with exact
LIT wavefunctions subjected to controlled multiplicative pointwise noise, the predicted stability
ordering is confirmed at every noise level over two orders of magnitude, with the
double-pole inversion reducing the noise-induced spread by a factor of $3$--$4$ relative
to standard practice.

Because the improvement enters mainly through post-processing, it can be adopted
immediately in existing LIT frameworks -- hyperspherical-harmonics, coupled-cluster, or
neural-network-based~\cite{LITReview,Bacca13,Sobczyk:2021dwm,Parnes:2026} -- wherever the
overlaps in Table~\ref{tab:kernels} can be evaluated. The practical payoff is largest
exactly where the LIT method is most strained: since inversion stability at fixed
$\Gamma$ improves several-fold, one may instead work at larger $\Gamma$ for a fixed
accuracy target, easing the solution of the LIT equation that limits current
applications to heavier systems.

\begin{acknowledgments}
This work was supported by the Israel Science Foundation under Grant No.\ ISF 2441/24.
\end{acknowledgments}

\bibliography{ProgramBib}

\end{document}